\documentclass[12pt,preprint]{aastex}
\def\dm15{${\rm \Delta m_{15}}$($B$)}
\def\kms{${\rm km\; s^{-1}}$}
\begin{document}

\title{The Luminosity of SN~1999by in NGC~2841 and
the Nature of the ``Peculiar'' Type~Ia Supernovae}
\author{Peter M. Garnavich\altaffilmark{1,}\altaffilmark{ 2}, 
Alceste Z. Bonanos\altaffilmark{1,}\altaffilmark{ 3}, 
Kevin Krisciunas\altaffilmark{2},
Saurabh Jha\altaffilmark{1},
Robert P. Kirshner\altaffilmark{1},
Eric M. Schlegel\altaffilmark{1}, 
Peter Challis\altaffilmark{1},
Lucas M. Macri\altaffilmark{1}
Kazuhito Hatano\altaffilmark{4,5},
David Branch\altaffilmark{4},
Gregory D. Bothun\altaffilmark{6}, and
Wendy L. Freedman\altaffilmark{7}}
\altaffiltext{1}{Harvard-Smithsonian Center for Astrophysics, 60 Garden Street, Cambridge, MA 02138}
\altaffiltext{2}{Physics Department, University of Notre Dame, Notre Dame, IN 46556}
\altaffiltext{3}{Wellesley College, Wellesley, MA 02481}
\altaffiltext{4}{Department of Physics and Astronomy, University of Oklahoma, Norman, OK 73019}
\altaffiltext{5}{Department of Astronomy and Research Center for the Early Universe, Univ. of Tokyo, Tokyo, Japan}
\altaffiltext{6}{Physics Department, University of Oregon, Eugene, OR 97403}
\altaffiltext{7}{Observatories of the Carnegie Institution of Washington, 813 Santa Barbara St., Pasadena, CA 91101} 

\email {pgarnavi@miranda.phys.nd.edu, kkrisciu@cygnus.phys.nd.edu \\
abonanos, kirshner, pchallis, eschlegel@cfa.harvard.edu \\
saurabh@astron.berkeley.edu \\
lmacri@noao.edu \\
branch@mail.nhn.ou.edu \\
nuts@bigmoo.uoregon.edu \\
wendy@ociw.edu 
}
\slugcomment{{\it The Astrophysical Journal}, volume 613, 2004 September
20}

\begin{abstract}

We present $UBVRIJHK$ photometry and optical spectroscopy of the so-called
``peculiar'' Type~Ia supernova 1999by in NGC~2841.  The observations began one
week before visual maximum light which is well-defined by daily observations. The
light curves and spectra are similar to those of the prototypical subluminous event
SN~1991bg. We find that maximum light in $B$ occurred on 1999 May 10.3 UT (JD
2,451,308.8$\pm$0.3) with $B=13.66 \pm 0.02$ and a color of
$B_{max}-V_{max}=0.51\pm 0.03$. The late-time color implies minimal dust
extinction from the host galaxy. Our photometry, when combined with the recent
Cepheid distance to NGC~2841 (Macri et al. 2001), gives a peak absolute magnitude
of $M_B=-17.15\pm 0.23$, making SN~1999by one of the least luminous Type~Ia
events ever observed. We estimate a decline rate parameter of
$\Delta$m$_{15}(B)=1.90$ mag, versus 1.93 for SN~1991bg, where 1.10 is typical for
so-called ``normal'' events. We compare SN~1999by with other subluminous events
and find that the $B_{max}-V_{max}$ color correlates strongly with the decline
rate and may be a more sensitive indicator of luminosity than the fading rate for
these objects.  We find a good correlation between luminosity and the depth of
the spectral feature at 580~nm, which {\em had} been attributed solely to Si~II.
We show that in cooler photospheres the 580~nm feature is dominated by Ti~II,
which provides a simple physical explanation for the correlation. Using only
subluminous Type~Ia supernovae we derive a Hubble parameter of
H$_0=75^{+12}_{-11}\; {\rm km\; s^{-1}\; Mpc^{-1}}$, consistent with values found
from brighter events.

\end{abstract}


\section{Introduction}

Type Ia supernovae (SNe Ia) are good distance indicators because they are intrinsically
bright and appear to have a relatively small dispersion in maximum brightness.
Phillips (1993) showed that a relation between the peak brightness and light curve
decline rate improves their utility as distance indicators. This
was exploited by Hamuy et al. (1996a), Riess, Press, \& Kirshner (1995) and Jha et al. (1999)
to measure the Hubble constant. Much more distant SNe Ia have been used to
determine the content of the universe (Garnavich et al. 1998; Schmidt et al. 1998; 
Riess et al. 1998, Perlmutter et al. 1999). 

SNe Ia are fairly homogeneous in spectral characteristics and intrinsic color at
maximum (Filippenko 1997), but in 1991 two spectroscopically ``peculiar'' SNe Ia
were discovered (Branch, Fisher, \& Nugent 1993). Near maximum light, SN~1991T
showed only a weak Si~II 615~nm line which is normally the strongest feature in
the optical band (Phillips et al. 1992; Filippenko et al. 1992a). It also displayed
a very slow light curve evolution compared to more typical events and was thought to
be more luminous than average. That same year SN~1991bg showed an extremely fast
light curve evolution as well as a red color at maximum light and strong Ti~II in
its spectrum (Filippenko et al. 1992b; Leibundgut et al. 1993). The maximum
brightness of SN~1991bg was also estimated to be two magnitudes fainter than normal
events. SN~1986G (Phillips et al. 1987) appears to have properties between normal
and the extreme case of SN~1991bg. Nugent et al. (1995) showed that the gross
spectral variations among all SNe Ia can be accounted for by simply varying the
photospheric temperature, which suggests peculiar events are just extreme tails of a
continuous distribution.
 
Since 1991, many more supernovae have been discovered serendipitously and in
systematic searches, and a handful of supernovae similar to SN~1991T and
SN~1991bg have been found and studied. Li et al. (2001) added SN~1999aa as a
peculiar sub-class which is similar to SN~1991T well before maximum light, but
with significant Ca~II absorption not seen in the original. From a volume-limited
sample, they find 36 percent of SNe Ia are somehow spectroscopically peculiar,
although Branch (2001) points out a selection bias that raises the fraction to 45
percent and casts doubt on the usefulness of the term ``peculiar''.

Only 16 percent of all the supernovae in the Li et al. sample were classified as
SN~1991bg-like with evidence of Ti~II in their spectra. Few such peculiar SNe Ia
have been studied in detail:  1992K (Hamuy et al. 1994), 1997cn (Turatto et al.
1998), 1998de (Modjaz et al. 2001), and 1999da (Krisciunas et al. 2001). Because of
their rapid evolution, they are often discovered after maximum brightness. Even 
rarer are SNe Ia that bridge the gap between so-called ``normal'' events and the
extreme SN1991bg-like explosions.  SN~1986G (Phillips et al. 1987) is one of these
intermediate events, but was highly reddened by dust in its host galaxy. More
recently, SN~1998bp (Jha et al. 1998; Jha 2002) and SN~2000bk (Krisciunas et 
al. 2001) appear to fall between the normal and extreme objects.

SN~1999by is another rare example of a ``peculiar'', fast-declining SN Ia. It was
discovered independently by R. Arbour, South Wonston, Hampshire, England, and by the
Lick Observatory Supernova Search (LOSS) on 1999 April 30 (Papenkova et al. 1999).  
SN~1999by is located in NGC~2841, an Sb galaxy, that has been host to three other
supernovae (SN~1912A, SN~1957A, and SN~1972R). An early spectrum by Gerardy \& Fesen
(1999) showed that SN~1999by was a Type~Ia event, and Garnavich et al. (1999) found
line ratios and Ti~II absorption consistent with a
SN~1991bg-like supernovae. SN~1999by is also one of the few
SNe Ia to show significant intrinsic polarization (Howell et al. 2001; see 
also Wang et al. 2003 regarding the case of SN~2001el).
 
Here, we present detailed photometric and spectroscopic observations of SN~1999by.
In addition to SN~1999by being fascinating as an unusual supernova, Hubble Space Telescope
($HST$) has studied Cepheid variables in its host galaxy (Macri et al.
2001), which makes SN~1999by useful for defining the distance scale.

\section{Observations}
\subsection{Photometry}

Optical broad band photometry of SN 1999by began six days after discovery with
the 1.2-m telescope at the Fred L. Whipple Observatory (FLWO). Twenty sets of
$UBVRI$ images were obtained between 1999 May~6 and June~21 (UT) using the
4-Shooter camera, which consists of four 2048$\times$2048 CCD chips. Images were
taken in a 2x2 binning mode with the supernova centered on chip~3. The image
scale was 0.64\arcsec\ pixel$^{-1}$. Late time $UBVRI$ images were obtained at FLWO
in 1999 November and December.

The images were bias corrected and flat fielded using the CCD reduction package
in IRAF\footnote[8]{Image Reduction and Analysis Facility, distributed by the
National Optical Astronomy Observatory, which is operated by the Association of
Universities for Research in Astronomy, Inc., under cooperative agreement with
the National Science Foundation.}.  We used the DAOPHOT package in IRAF to obtain
instrumental magnitudes of the supernova and 10 local standard stars using
aperture photometry. Tests using point-spread-function (PSF) fitting photometry
showed no significant difference between the derived magnitudes. SN~1999by appeared
about 3\arcmin\ from the center of NGC~2841 where there is still significant
light contributed by the disk. However, the galaxy is relatively smooth at the
resolution of FLWO with only a mild light gradient at the supernova, permitting
an accurate local background to be estimated from an annulus around the star. At
late times, no contamination from faint stellar sources is seen at the position
of the supernova.

\begin{figure}[h]
%\plotone {f1.ps}
\plotfiddle{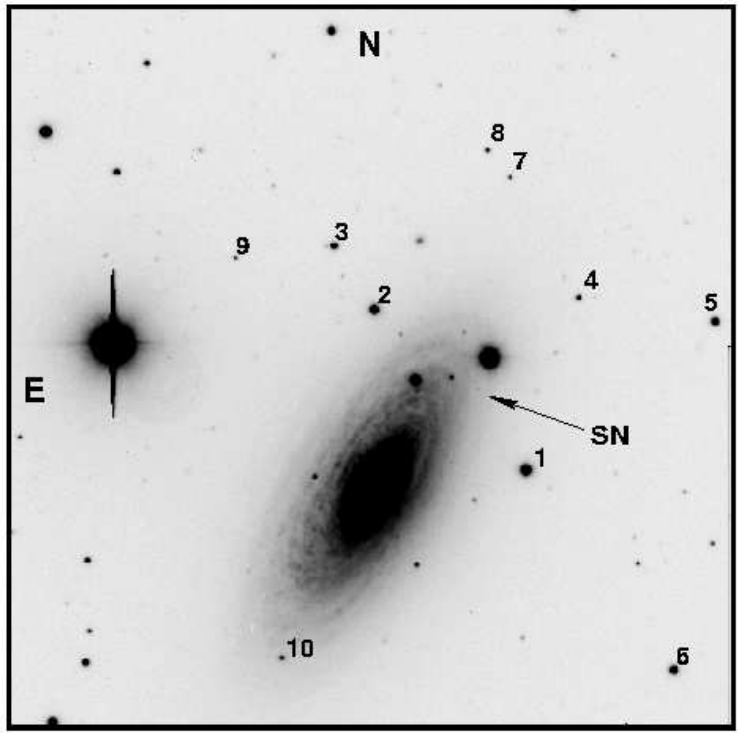}{5.5cm}{0}{100.0}{100.0}{60.0}{-100.0}
\caption{The field around NGC~2841 with local comparison stars marked. The field
of view is 10.9 by 10.9 arcminutes. The image was taken in 1999 December
well after the supernova maximum. The supernova location was at the
tip of the arrow.}
\label{fig:standards}
\end{figure}

\begin{figure}[h]
\epsscale{0.7}
\plotone {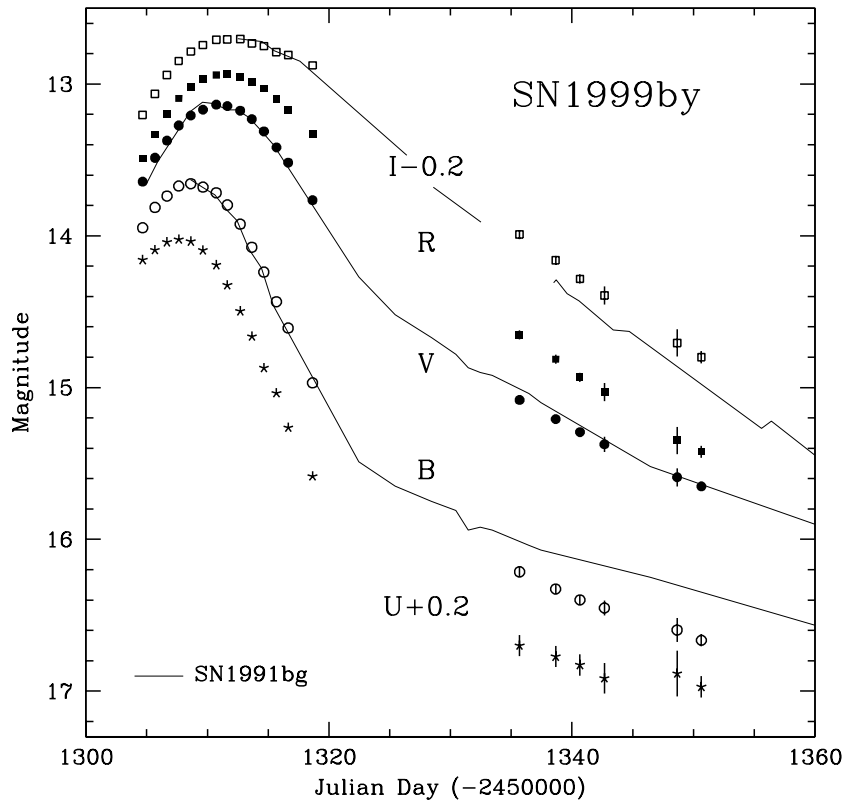}
\caption{Light curves of SN 1999by in the U, B, V, R and I filters.
The solid lines show the light curves of SN 1991bg from Leibundgut et al. (1993),
with the curves shifted in magnitude and time to match the light curves
of SN~1999by at maximum.  For the data near maximum, the 
uncertainties of the photometry are on the order of the size of the points.}
\label{fig:lightcurve}
\end{figure}

Local standard stars were calibrated on four photometric nights
at FLWO and two photometric nights at the Vatican Advanced Technology
Telescope (VATT). On all those nights Landolt standard stars (Landolt 1992)
were observed over a range of airmasses allowing for the derivation of 
linear extinction and color coefficients. The standard magnitudes derived for
the local calibrators are given in Table~1 corresponding to the stars marked
in the finder chart given in Figure~\ref{fig:standards}. For local calibrators 
observed on multiple nights, the root-mean-square (rms) scatter was 
0.03 mag.\footnote[9]{For secondary standards we prefer to present the
mean magnitudes to a resolution of 0.001 mag. In our experience secondary
standards have uncertainties of several thousandths of a magnitude to
a few hundredths, depending on the filter.} The 
supernova instrumental magnitudes from May and June were converted to standard 
magnitudes using local standards stars 1 to 4. The resulting $UBVRI$
photometry is provided in Table~2 and displayed in Figure~\ref{fig:lightcurve}.
Our standard star calibration has substantially improved over that used by
Bonanos et al. (1999) which was based on a single non-photometric night,
and our $V$ maximum is now consistent with that found by Toth \& Szab\'o (2000).

The optical light curves of SN~1999by shown in Figure~\ref{fig:lightcurve} are
similar to those of the peculiar Type Ia SN~1991bg in their rate of rise and decline
from maximum.  Although our coverage 10 to 30 days after $B_{max}$ is poor, the
SN~1999by light curves in the $R$ and $I$ bands are consistent with those of
SN~1991bg, which lacked the prominent secondary maximum seen in Type Ia SNe of
mid-range decline rates.

From the highest quality image we derive an accurate position for the supernova of
$\alpha$ = 9:21:52.07, $\delta$ = 51\arcdeg\ 00\arcmin\ 06.54\arcsec\ (2000) based
on the $HST$ Guide Star Catalog coordinates. Using 14 stars in the field of NGC~2841
for the plate solution we get a scatter of 0.2$''$ (rms) which represents the
expected accuracy of the astrometry. This is consistent with the position given by
the discoverers (Papenkova et al. 1999).

\begin{figure}[h]
\epsscale{0.7}
\plotone{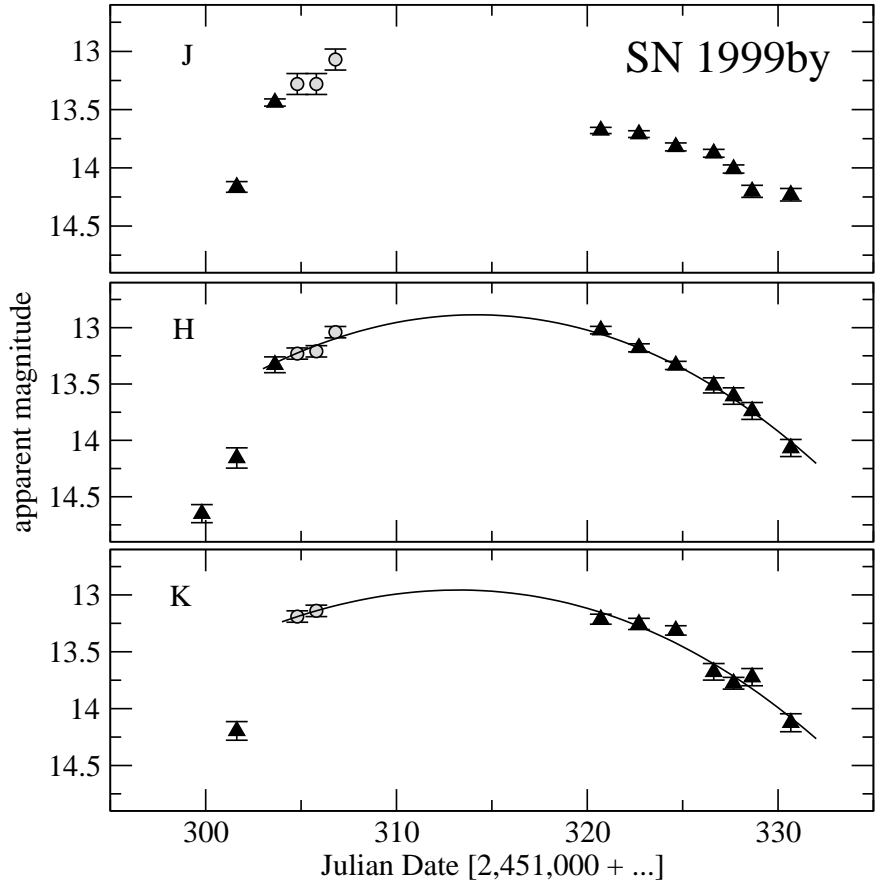}
\caption{$JHK$ light curves of SN 1999by.  The open circles
are data from H\"{o}flich et al. (2002), while the triangles are
data from the FLWO 1.2-m telescope given in Table 3. The
solid lines show a polynomial fit to the points to aid in
estimating the position of maximum light.}
\label{fig:jhkdat}
\end{figure}

In Table 3 we give near infrared (IR) data obtained with the FLWO 1.2-m
telescope using STELIRCAM, an imaging instrument containing two InSb
arrays of 256 $\times$ 256 pixels which allows simultaneous blue channel 
(e.g. $J$-band) and red channel (e.g. $H$- or $K$-band) data acquisition.
Data were taken at a resolution of 1.2 \arcsec\ pixel$^{-1}$.  On photometric
nights bright infrared standards of Elias et al. (1982) or fainter
standards of Persson et al. (1998) were observed.  From four photometric
nights we find $J$ = 12.100 $\pm$ 0.013, $H$ = 11.732 $\pm$ 0.054, and 
$K$ = 11.727 $\pm$ 0.033 for field star number~1.  These values are to
be compared with $J$ = 12.099 $\pm$ 0.028, $H$ = 11.728 $\pm$ 0.029,
and $K$ = 11.632 $\pm$ 0.027 from the Two Micron All Sky Survey (2MASS).
Thus, there is excellent agreement of the IR zeropoints in $J$ and $H$,
but a significant difference of 0.1 mag between our 
$K$-band photometry and the 2MASS value for field star number~1. Krisciunas et al. (2004b)
has found an offset and scatter between the 2MASS $K$-band magnitudes and the
same stars calibrated with Persson et al. (1998) standards which can account for
this inconsistency.
In Figure~\ref{fig:jhkdat} we show our $JHK$ photometry of SN~1999by
along with three nights of data given by H\"{o}flich et al. (2002).  

\subsection{Spectroscopy}

Spectroscopic observations of SN 1999by were made using the FLWO 1.5-m telescope
with the FAST spectrograph using a 300 line mm$^{-1}$ grating and a 3\arcsec\
wide slit (Fabricant et al. 1998). A majority of the spectra were taken with a single
grating setting covering 362~nm to 754~nm. On two nights the observations were
done at two grating tilts providing wavelength coverage from 327~nm to just over
900~nm and three spectra were obtained with a 600 line mm$^{-1}$ grating giving twice the
resolution but half the typical coverage.  Interference fringes were a
severe problem at wavelengths longer than 780~nm and could not be completely
removed from the data. A total of 18 spectroscopic observations was made between
1999 May~6 and June~22 UT. A log of the observations is given in Table~4.

\begin{figure}[h]
\epsscale{0.6}
\plotone {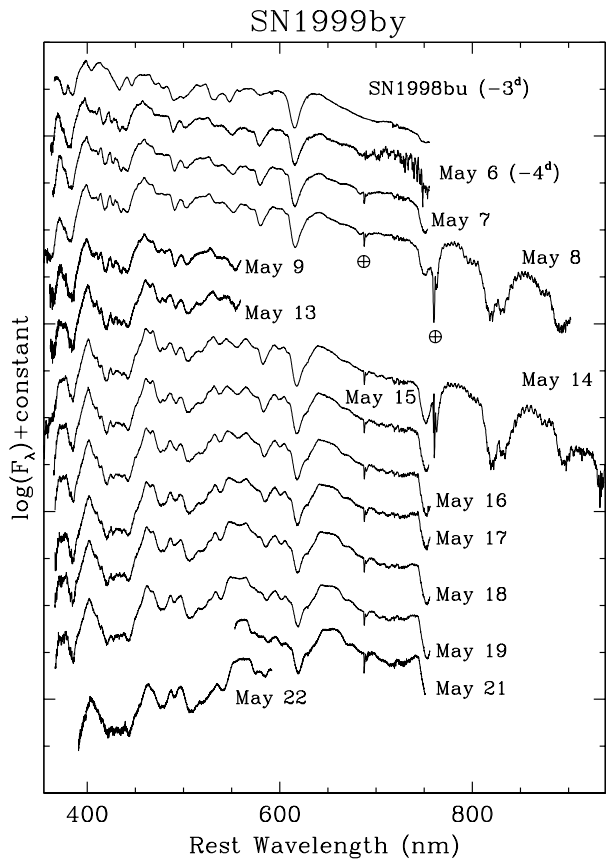}
\caption{Spectra of SN 1999by in May, 1999 from the FLWO 1.5m telescope.
For comparison we show the normal but dust reddened SN1998bu three days
before maximum light.}
\label{fig:mayspec}
\end{figure}

\begin{figure}[h]
\epsscale{0.6}
\plotone {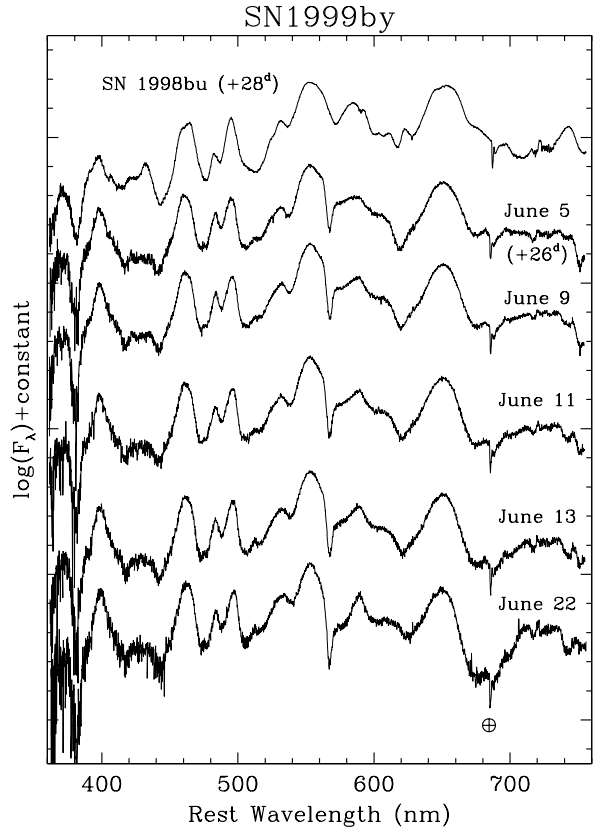}
\caption{Spectra of SN 1999by in June, 1999 during the early nebular stage.
For comparison, the top spectrum is that of SN~1998bu at an age of $+28$ days
after $B_{max}$.  The narrow absorption feature seen in SN~1999by near 570~nm is identified as Na~I.}
\label{fig:junespec}
\end{figure}

We reduced the spectra using IRAF. The two dimensional CCD exposures were bias
and dark corrected and then flat fielded. Using the APEXTRACT package, we
extracted the 1D spectrum at the supernova position, subtracting the sky
estimated along two strips parallel to the spectrum. The supernova signal was
strong enough that it was used to trace the centroid along the dispersion.
Exposures of a HeNeAr lamp taken after each supernova observation were used to
calibrate the wavelength. Flux calibration was done by using the exposures of
spectroscopic standard stars taken during the night. The spectrograph slit was
set at the parallactic angle for both the supernova and the standard star
observations to reduce differential slit losses. Not all the observations were
made under photometric conditions.  Finally, the data were corrected for the
Doppler shift due to the radial velocity of NGC 2841 (638
\kms).\footnote[10]{This is the heliocentric radial velocity of the host galaxy
from the NASA/IPAC Extragalactic Database (NED). As noted by Vink\'o et al.
(2001), the radial velocity of SN~1999by would contain a 300 \kms uncertainty due
to the rotation of the host galaxy.} The reduced spectra are shown in
Figures~\ref{fig:mayspec} and \ref{fig:junespec}.

\section{Discussion}
\subsection{The Light Curve}

The date and magnitude of maximum light in all the filters were estimated by fitting
second and third degree polynomials to the May data. Because of the daily sampling,
the $UBVRI$ curves are very well defined and the scatter about the fits was 0.01 mag
or less. The results are given in Table~5. A two day difference between the times of
$B$ and $V$ maximum (T($B_{max}$) and T($V_{max}$)
is typical of SNe Ia (Leibundgut 1988) and appears from
SN~1999by to be true for fast-declining events as well.  The Galactic latitude of
NGC~2841 is $+44^\circ$ which means that the extinction through our galaxy is low
but not necessarily insignificant. The reddening estimate from Schlegel, Finkbeiner,
\& Davis (1998) is E$(B-V)=0.016$ mag, and this correction has been applied to the
maximum magnitudes given in Table~6 assuming the standard Cardelli et al. (1989)
extinction law.

\begin{figure}[h]
\epsscale{0.7}
\plotone {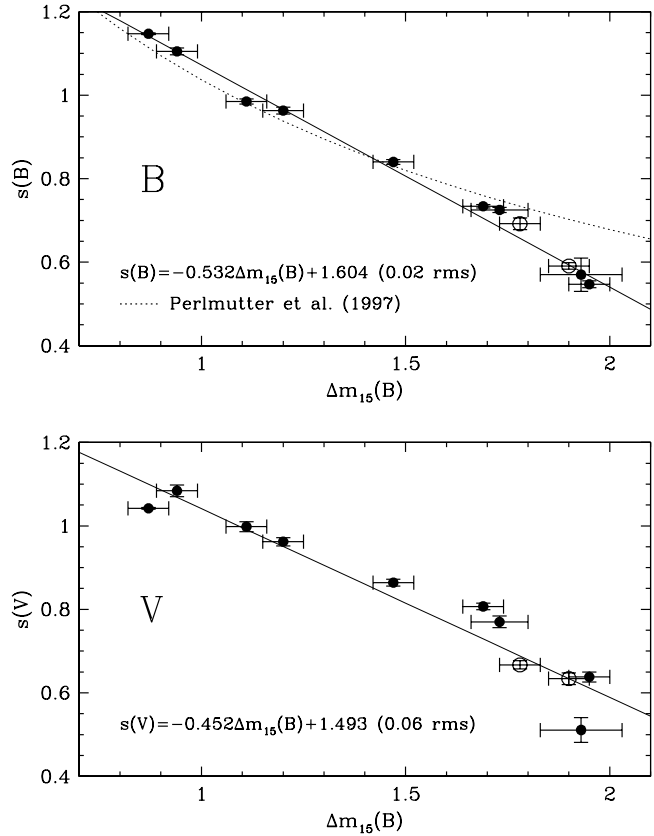}
\caption{The relationship between the stretch parameter, $s$, and \dm15. The
solid points show the \dm15\ standard SNe Ia from Hamuy et al. (1996) plus
SN~1998de from Modjaz et al. (2001). The open points show SN~1998bp and the
position of SN~1999by for its measured stretch. The solid line is the best fit
line through the solid points. The dotted line is the conversion of a stretch
parameter to \dm15\ from Perlmutter et al. (1997) where the light curve templates
were restricted to \dm15$<1.75$.}
\label{fig:dm15tos}
\end{figure}

The number of magnitudes a supernova fades during the 15 days after maximum
brightness, \dm15, is a popular way to parameterize a light curve shape. Directly
measuring this parameter is difficult from our data because of the lunar gap
starting ten days after $B_{max}$. It is also unwise to use the standard
template-fitting technique when there are no known SNe Ia significantly faster
than SN~1991bg. Instead, we apply a light curve ``stretch method'' developed for
high-redshift supernovae by Perlmutter et al. (1997). In our implementation, the
time axis of the Leibundgut templates (Leibundgut 1988), restricted to $-5<{\rm
days}<15$, are multiplied by a parameter, $s$, which compresses or expands the
template light curve in the time domain, mimicking the variation in decline rate.  
A $\chi^2$ minimization is applied to an observed light curve after correcting
for time dilation, where the stretch factor, time of maximum and magnitude of
maximum are free parameters. We then applied this technique to the \dm15\
standard templates defined by Hamuy et al. (1996a) as a way to calibrate the
stretch factor against \dm15. Because we are interested in the fast-declining end
of the distribution, we added the well-observed event SN~1998de with \dm15
$=1.95\pm 0.05$ (Modjaz et al. 2001) as a standard light curve.

\begin{figure}[h]
\epsscale{0.7}
\plotone {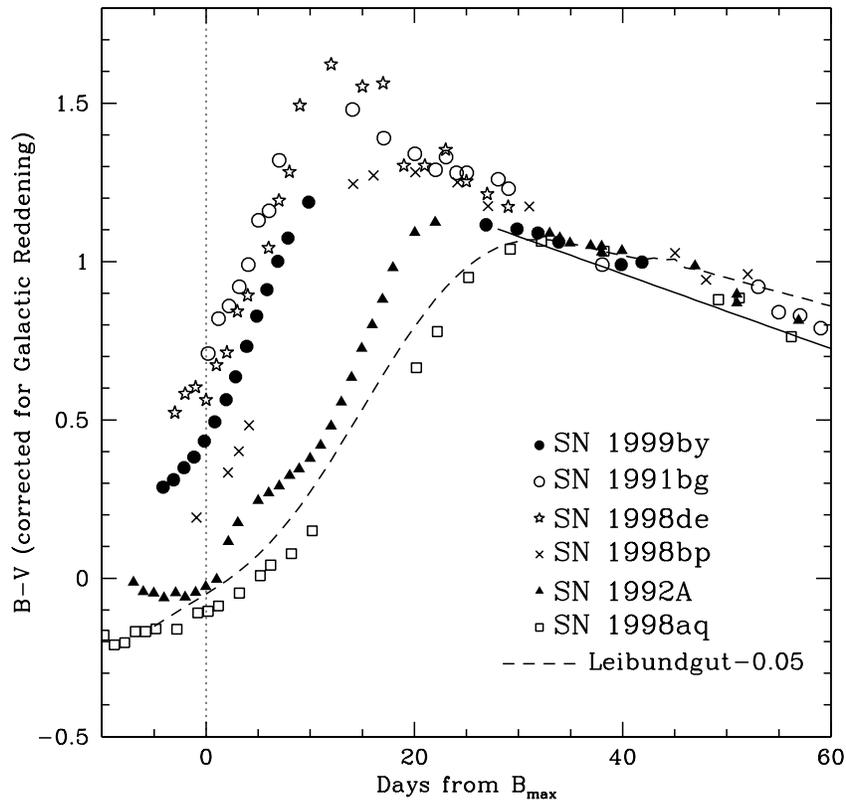}
\caption{$B-V$ color for seven SNIa corrected for Galactic extinction. 
The solid line is the empirical relation of Lira (1995) for unextinguished
supernovae at late-times. The Leibundgut templates have been shifted
by 0.05 mag to correct for the average reddening to SNIa from
Phillips et al. (1999).}
\label{fig:color}
\end{figure}

The calibration between the stretch parameters for $B$ and $V$ and \dm15\ are shown
in Figure~\ref{fig:dm15tos}. The conversion of stretch to \dm15\ works best for the
$B$-band with a scatter of only 0.04 mag in \dm15. Stretch applied to the $V$-band
gives a larger scatter, but consistent results.  Direct measurement of the \dm15\
parameter should become slightly non-linear for these very fast-declining events
since the first inflection point in the light curve, which signals the onset of the
nebular phase, shifts to within 15 days of $B_{max}$.  This likely contributes to
the scatter in Figure~\ref{fig:dm15tos} but does not have a large effect on our goal
of estimating \dm15\ for SN~1999by.

We estimate SN~1999by to have \dm15$=1.90\pm 0.05$ mag, making it a slightly
slower declining light curve relative to SN~1991bg and SN~1998de, but still one
of the fastest fading Type Ia light curves ever observed. For SN~1998bp (Jha
2002) we directly measure \dm15$=1.78\pm 0.05$, consistent with the derived
stretch parameters.

The observed $B_{max}-V_{max}$ color for SN~1999by is 0.51$\pm 0.03$ mag.  This
is extremely red compared to SNe Ia of mid-range decline rates (Phillips et al.
1999), which have $B_{max}-V_{max} < 0.1$. Since the host of SN~1999by is a
spiral with obvious dust patches, reddening may explain this color. However,
other fast declining SNe Ia appear to be intrinsically red (Leibundgut et al.
1993; Modjaz et al. 2001) near maximum and many of these occurred in elliptical
hosts with minimal dust. Figure~\ref{fig:color} plots the color curves for
SN~1999by compared with other fast-declining events.

\begin{figure}[!h]
\epsscale{0.7}
\plotone {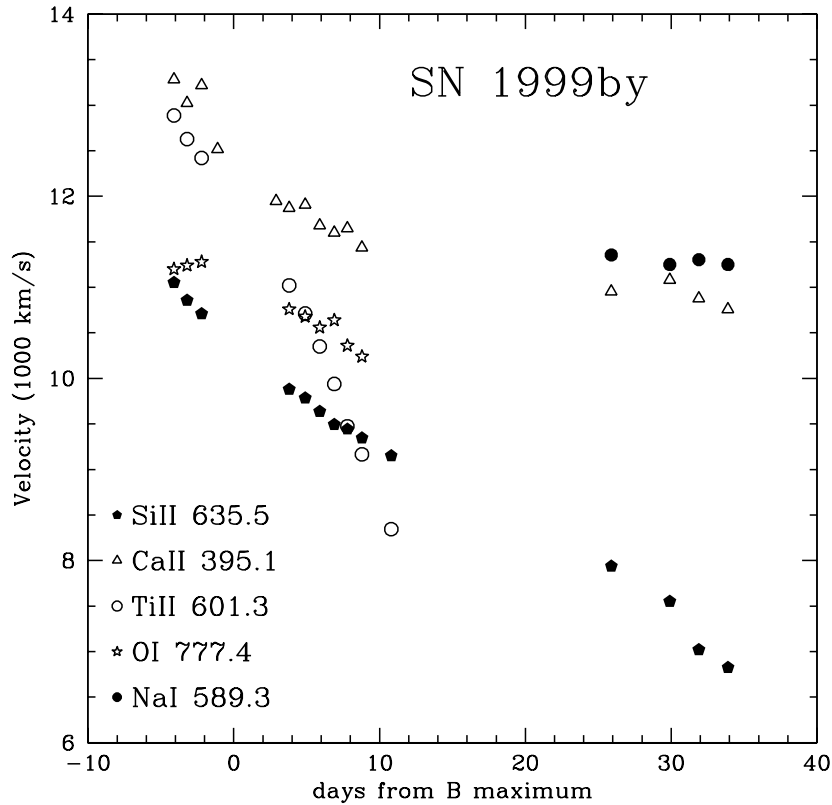}
\caption{Absorption minimum velocities of Si~II 635.5~nm, Ca~II 395.1~nm,
Ti~II 601.3~nm, O~I 777.4~nm and
Na~I 589.3~nm lines in SN~1999by.}
\label{fig:velocity}
\end{figure}
  
Lira (1995) has shown that SNe Ia with low extinction tend to have uniform $B-V$
colors between 30 and 90 days after the time of $V$-band maximum. This appears to
hold for fast decliners as well, although it has been tested with very few events.
This fact has been exploited by Phillips et al. (1999) to correct the brightness of
a large sample of SNe Ia for dust extinction. The Lira relation, \begin{equation}
(B-V)_{o} = 0.725 - 0.0118 ({\rm T}({V_{max}})-60) \end{equation} is also plotted in
Figure~\ref{fig:color}. 
%The late-time color of SN~1999by is consistent with fast decliners in ellipticals, 
%suggesting minimal host extinction. 
The average difference between the Lira relation and the late-time supernova color 
is only 0.014$\pm 0.020$ mag, suggesting that extinction from the host galaxy is 
minimal and may indicate that the supernova exploded on the near side of the 
NGC~2841 disk.

Though a combination of our IR photometry and that of H\"{o}flich et al. (2002)
shows a gap at the time of maximum light, we can fit a polynomial to the $H$- and
$K$-band data to derive the maximum magnitudes in those two bands (see Figure~\ref{fig:jhkdat}).
We do not feel that the $J$-band photometry gives us the same
option.  The $H$- and $K$-band maxima occurred about 5 days {\em after} T($B_{max}$).
All other known Type Ia supernovae, including the rather fast decliner
SN~1986G, have IR maxima that occurred about 3 days {\em before} T($B_{max}$)
(Meikle 2000, Krisciunas et al. 2004b).

\subsection{The Spectra}

The first spectrum was taken 4 days before T($B_{max}$) and the data
follow the spectral development without a major interruption until 12 days after
maximum. 

\begin{figure}[!h]
\epsscale{0.7}
\plotone {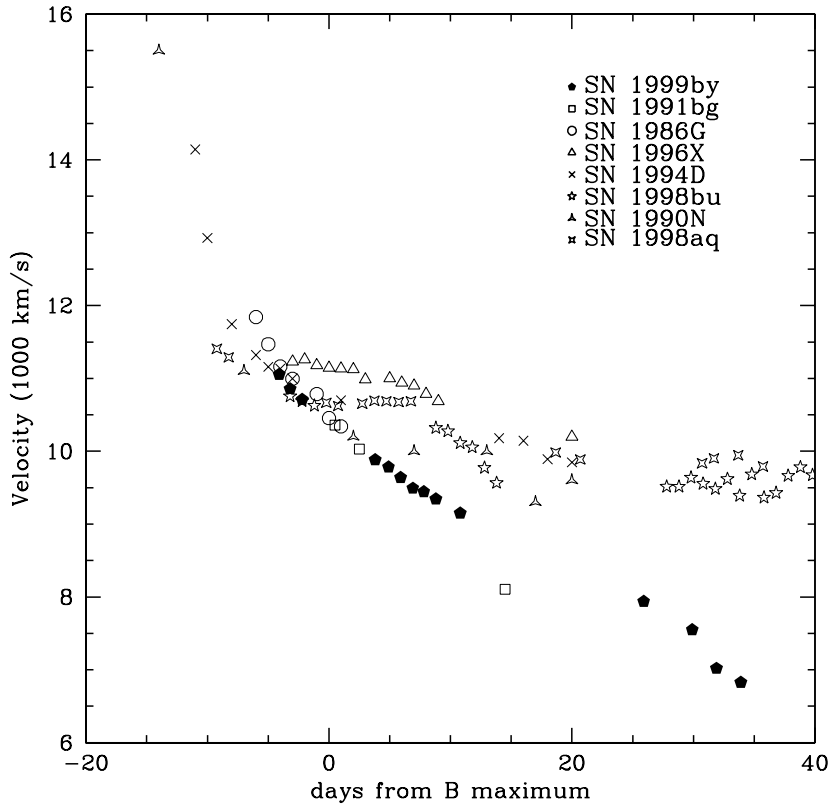}
\caption{Comparison of the Si~II 635.5~nm absorption velocity for SN~1999by and
other SNIa. While the velocities are similar at maximum, the subluminous
supernovae continue to decline in photospheric velocity so at late-times
they are distinguishable from more luminous events.}
\label{fig:si2}
\end{figure}

Comparing the spectrum of SN~1999by at maximum light to spectra of SNe Ia of
mid-range decline rates, we see the characteristic Si~II feature at 635.5 nm is
blueshifted to $\approx$ 615~nm. There is also a prominent absorption band between
420~nm and 440~nm which is unique to fast-declining SNe Ia.  The band as well as
absorption at 470~nm and 505~nm were observed in SN~1991bg and have been attributed
to Ti~II (Filippenko et al. 1992a). The 420~nm band deepens over the observing span
but was present even before maximum.

Another feature unique to fast-declining events is the deep O~I 777.4~nm line
that appears near the edge of the spectral range on most nights. We plot the
velocity of O~I as measured by the minimum of the absorption trough in
Figure~\ref{fig:velocity} along with the velocities derived for other prominent
lines such as Si~II 635.5~nm, Ca~II 395.1~nm and at late times Na~I 589.3~nm.  
The absorption-line velocities have been adjusted for the recession velocity of NGC~2841 and
because the photospheric expansion is a significant fraction of the speed of light,
a relativistic correction has been applied.  Mazzali et al. (1997) found that the O~I
line in SN~1991bg
remained constant at 11000~\kms\ for two weeks after maximum light. In
SN~1999by the O~I line does decline in velocity from 11200~\kms\ before
maximum to 10200~\kms\ nine days after maximum, which is not as steep as for
Si~II. This may mean the oxygen shell in SN~1999by is deeper than in SN~1991bg.

\begin{figure}[!h]
\epsscale{0.7}
\plotone {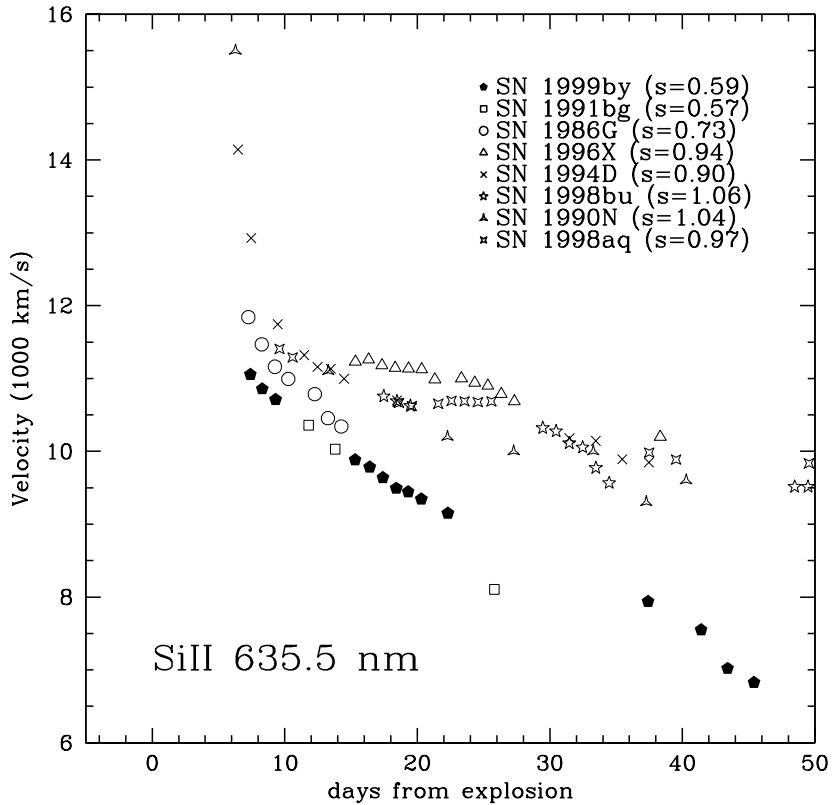}
\caption{Comparison of the Si~II 635.5~nm absorption velocity for SN~1999by and
other SNIa relative to the time of the explosion. The time of the explosion
is estimated from the stretch parameter.  }
\label{fig:si2_exp}
\end{figure}

The velocity of the Si~II 635.5~nm absorption versus time is compared with a
range of other SNe Ia in Figure~\ref{fig:si2}.  The velocities of the other SNe were
taken from Jha et al. (1999) and Jha (2002).  The Si~II velocity before maximum
light is similar to SNe Ia of mid-range decline rates, but SN~1999by shows a
continual drop in velocity like that of SN~1986G, while the typical SN Ia levels
off at between 11000~\kms\ and 10000~\kms. A steep decline in velocity was also
found by Vink\'o et al. (2001) for SN~1999by.  We find that the average slope
around maximum is 130~\kms day$^{-1}$ while most so-called normal events have
flatter slopes.  Only SN~1991bg, with a slope of 160~\kms d$^{-1}$, appears to
decline faster. Wells et al. (1994) noted a correlation between \dm15\ and Si~II
velocity decline rate. These data at least show that Type Ia supernovae with the
fastest declining light curves also have the fastest declining Si~II velocities
after T($B_{max}$).

Comparing the spectra of fast-declining SNe Ia with more typical events using
the time of $B_{max}$ as the reference is not ideal. It might be better to
compare the velocities observed in various supernovae using the time of
explosion as the zero point. Riess et al. (1999b) find that for a nominal SN Ia
(\dm15 =1.1) the interval from explosion to T($B_{max}$) is 19.5$\pm 0.2$ days,
and this is consistent with SNe Ia observed at high redshift when a stretch
correction is applied based on the light curve shape (Goldhaber et al. 2001).
It is therefore possible to approximate the day of the explosion as
T($exp$)=T($B_{max}$)$-19.5\times s$, where $s$ is the light curve stretch
parameter based on the Leibundgut templates. Since $s$ varies by only about
10 percent for normal luminousity supernovae, this provides only a day or two shift in the
relative times for the observed velocities. But in SN~1991bg-like events the
time of maximum can correspond to as little as 11 days after explosion. In
Figure~\ref{fig:si2_exp} we plot the observed velocities of the Si~II
absorption in several supernovae versus the time of explosion estimated as
above. When the date of explosion is used as the zero point of the Si~II
velocities, SN~1999by and SN~1991bg are always lower than normal events.
Also, the velocities of intermediate supernovae, such as SN~1986G, fall between
the extreme fast-decliners and the normal supernovae. For fast-declining
events, there appears to be a simple correlation between the light curve shape
and photospheric velocity when measured relative to the time of explosion.

\begin{figure}[!h]
\epsscale{0.8}
\vspace{-3.0cm}
\plotone {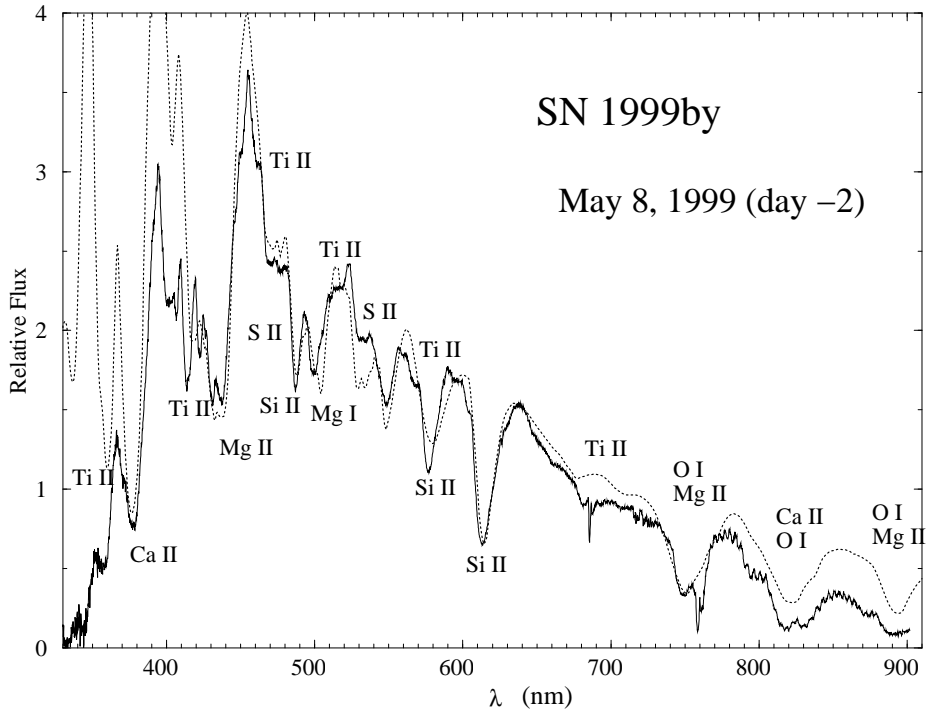}
\vspace{-4.3cm}
\caption{A model fit to SN~1999by two days
before $B_{max}$. The solid line is the observed data and the dotted line is the
model. Mg~I near 500~nm is newly identified in a SNIa.}
\label{fig:model07}
\end{figure}

The deep, narrow absorption observed at 567.5~nm in the June spectra (+25 to +35
days) was seen in SN~1991bg and attributed to Na~I 589.3~nm by Filippenko et al.
(1992b). But models by Mazzali et al. (1997) can not reproduce this feature
without adding extra sodium and customizing the distribution to match the small
velocity width. If the line is attributed to sodium, then the velocity
in SN~1999by is a constant 11300~\kms\ compared to 10500~\kms\ in SN~1991bg
(Turatto et al. 1996) and its width is only 1200~\kms .

\begin{figure}[!h]
\epsscale{0.8}
\vspace{-3.0cm}
\plotone {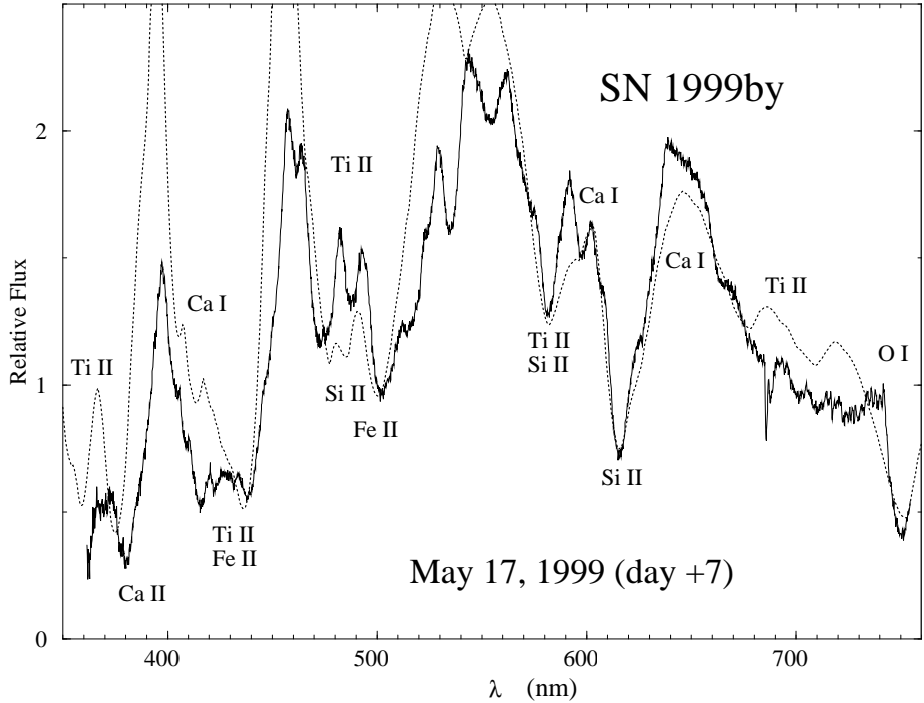}
\vspace{-4.3cm}
\caption{A model fit to the observed spectrum of SN~1999by at seven days
after $B_{max}$. The solid line is the observed data and the dotted line is the
model. Ca~I is identified at a number of locations in this early nebular phase
spectrum.}
\label{fig:model16}
\end{figure}

We fit models to the SN~1999by spectra using the parameterized spectral
synthesis code SYNOW (Fisher et al. 1999; Hatano et al. 1999a).
Figure~\ref{fig:model07} shows the fit to the May~8 (UT) spectrum taken two
days before T($B_{max}$). The fit longward of 420~nm is excellent and shows that
most of the features come from Si~II, Mg~II, Ca~II, O~I and Ti~II. We match the
absorption feature near 500~nm with Mg~I, which has not been identified in a
SN Ia before.  The fit to the May~17 (+7 d) spectrum
(Figure~\ref{fig:model16}) shows the Ti~II has increased in strength in the
blue and the temperature is low enough to allow neutral calcium to appear in
the spectrum.

\begin{figure}[!h]
\epsscale{0.8}
\vspace{-3.0cm}
\plotone {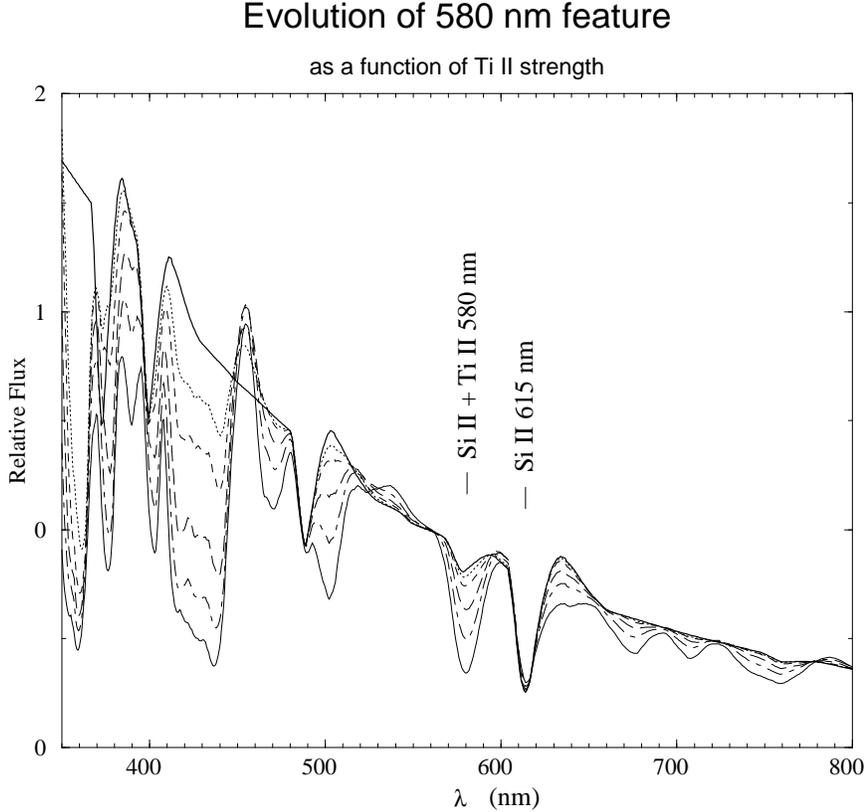}
\vspace{-4.3cm}
\caption{A SYNOW model showing only lines of Si~II and Ti~II on a 12000~K
continuum. The thick solid line shows the Si~II absorption spectrum with no
contribution  from Ti~II. The thin lines show how the spectrum is modified
by varying the optical depth of Ti~II. Note the feature at 580~nm, which is normally
dominated by Si~II, has a large contribution from Ti~II when the Ti~II
abundance is significant. }
\label{fig:tau}
\end{figure}

\begin{figure}[!h]
\epsscale{0.7}
\vspace{-2.0cm}
\plotone {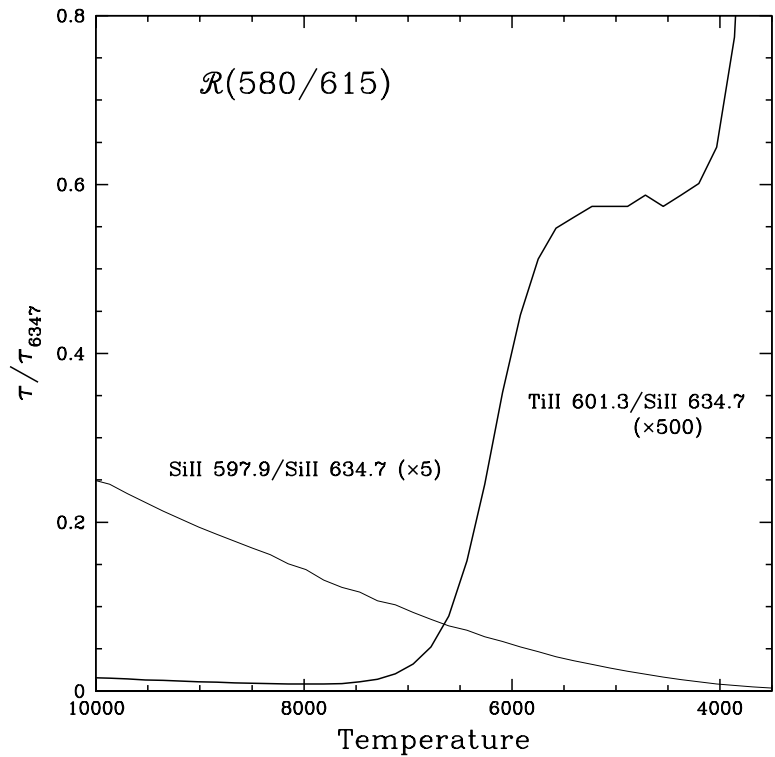}
\caption{Ratio of the optical depth of the Si~II line 597.9~nm and Ti~II
line 601.3~nm to the optical depth of Si~II 634.7~nm for a range of
temperatures. The optical depths have been multipled to approximate
the observed 580/615 depth ratio. The Si~II 597.9 strength is expected to
decline relative to Si~II 634.7 with decreasing temperature while Ti~II
begins an exponential rise at 7000~K.}
\label{fig:tauratio}
\end{figure}

A great advantage of SYNOW is that the strength of a single ion can be
varied to determine its effect on the entire spectrum. Since Ti~II is
such an important contributor to the spectra of fast declining SNe Ia, we
plot the variation in Ti~II optical depth with a fixed Si~II optical
depth in Figure~\ref{fig:tau}. The continuum temperature is
12000 K for all the models.  The blue region of the spectrum,
especially the 400-440~nm band is strongly absorbed by the Ti~II.
Surprisingly, an absorption feature at 580~nm appears at high Ti~II
optical depths. This has commonly been attributed to Si~II (e.g.
Filippenko 1997) and, indeed it is dominated by Si~II 597.9~nm when the
Ti~II contribution is small. However at low temperatures a number of
Ti~II lines take over. This nicely explains the correlation between the
580~nm line depth and \dm15\ found by Nugent et al. (1995). They noted a
difficulty in physically accounting for the observed increase of the
580~nm to 615~nm depth ratio with decreasing temperature since the ratio
of the two lines should mildly decrease with lowering temperature.  This
is confirmed in Figure~\ref{fig:tauratio} which shows the optical depth
ratio $\tau$(597.9)/$\tau$(634.7) versus temperature from the study of
line strengths in supernova atmospheres by Hatano et al. (1999b).  The
optical depth ratio of the representative Ti~II 601.3~nm to Si~II
634.7~nm line is also shown and demonstrates a very rapid increase with a
small temperature drop. We conclude that the 580~nm depth to 615~nm depth
ratio, called ${\mathcal R}$(Si~II) by Nugent et al. (1995) is dominated
by Si~II at high temperatures and Ti~II at low temperatures.  The rise of
the Ti~II lines appears to be very rapid as the optical depth relative to
Si~II increases exponentially with decreasing temperature. This suggests
that the so-called ``peculiar'' SNe Ia may represent a rather small
deviation from normal events but with a rapid line blanketing by Ti~II
contributing to the red color, the fast $B$-band decline, the faint
$B$-band luminosity and the unusual spectral 
characteristics. The key physical parameter controlling the photospheric temperature
may be the $^{56}$Ni yield.  SNe Ia
are generally considered explosions of CO white dwarfs which have
reached the Chandrasekhar limit from the transfer of mass from a
nearby donor star.  Contardo, Leibundgut, \& Vacca (2000) found that
the $^{56}$Ni yield of SN~1991bg was 8 times less than the brightest
object in their sample, SN~1992bc.

\begin{figure}[!h]
\epsscale{0.7}
\vspace{-2.0cm}
\plotone {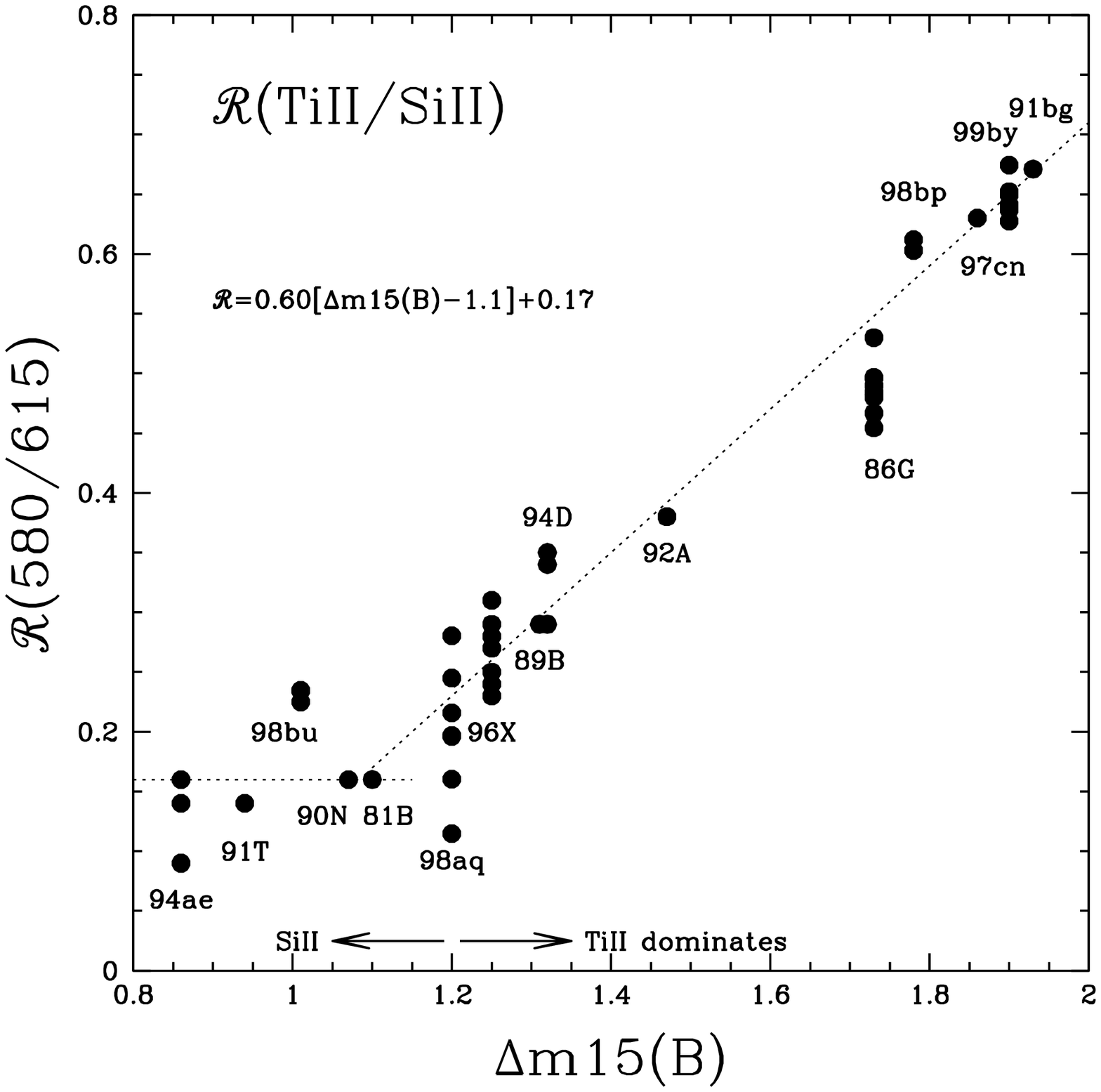}
\caption{Ratio of 580~nm to 615~nm line depth for 15 supernovae. Each point
represents an individual spectrum and the data is restricted to
$\pm$3 days from $B_{max}$. The fit applies to supernovae with \dm15$>1.2$.
For \dm15$<1.2$ the line ratio is dominated by Si~II and is expected
to be nearly constant, making the ratio a poor
indicator of luminosity, decline rate or temperature.}
\label{fig:ratio}
\end{figure}

The 580/615~nm line depth ratio, which we will now call ${\mathcal
R}$(Ti~II/Si~II), should still be a useful indicator of decline rate (and
intrinsic brightness), at least for the low temperature events.  In
Figure~\ref{fig:ratio} we plot ${\mathcal R}$(Ti~II/Si~II) for supernovae with a
wide range of the \dm15\ parameter, restricting the spectra to within three days
of $B_{max}$. We expect the ratio to be relatively flat with temperature when the
580~nm feature is dominated by Si~II and begins to increase when Ti~II is
present. The rise appears to begin for \dm15$>1.2$, so Ti~II is present in the
red end of the spectrum even in more slowly declining, but otherwise normal SNe
Ia.  The strong Ti~II bands in the blue are probably not apparent for
$1.2<$\dm15$<1.7$ because the continuum optical depth is much higher there than
in the red, but more detailed modelling of this effect is required to confirm
this conjecture.

\section{Comparison with other SNe Ia: What do we mean by ``peculiar''?}

\subsection{Color}

\begin{figure}[!h]
\vspace{-1.0cm}
\epsscale{0.7}
\plotone {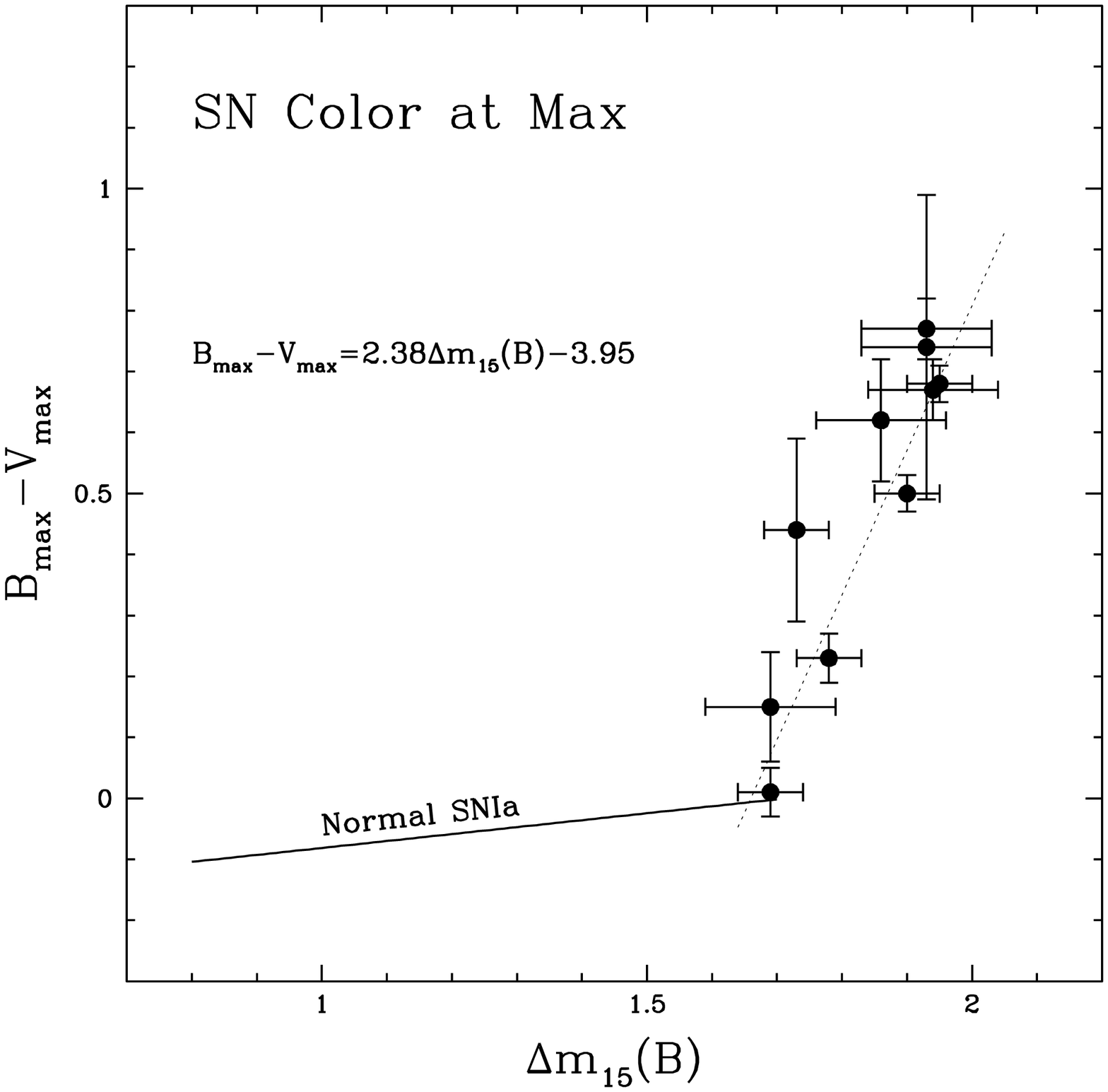}
\caption{The $B_{max}-V_{max}$ color for SNIa. The solid line shows
the color derived by Phillips et al. (1999) for normal SNIa. The dotted
line is the weighted linear fit to the supernovae listed in Table~6.}
\label{fig:dm15color}
\end{figure}

So-called ``peculiar'' SNe Ia with fast-declining light curves are defined
spectroscopically by the presence of strong Ti~II absorption (Branch, Fisher, \&
Nugent 1993). But we have shown that Ti~II is detectable for \dm15$>1.2$, and its
strength continuously increases between the so-called ``normal'' events and the
extreme SN~1991bg-like objects. The color of the supernova at maximum may be the
best discriminator. Phillips et al. (1999) has shown that the intrinsic
$B_{max}-V_{max}$ color of SNe Ia is less than 0.1 for \dm15$<1.7$. A
compilation of fast-declining SNe Ia is given in Table~6 for events with
\dm15$\geq1.69$. We have left out SN~1992br (\dm15=1.69; Hamuy et al. 1996b) and
SN~1996bk (\dm15=1.75; Riess et al. 1999a) because their light curves are poorly
defined. SN~1992K was not discovered until an estimated 12 days after maximum and
its parameters are poorly determined but we include it here because it is well
established in the literature.  The $B_{max}-V_{max}$ color versus \dm15\ for the
objects with fast light curves is shown in Figure~\ref{fig:dm15color} and
demonstrates a sharp break near \dm15$\sim 1.7$.  Despite the rapid change in
slope, the fast-declining events connect well to the end of the ``normal'' color
distribution. No events with \dm15$>1.7$ have been found with blue colors, so
there remains a monotonic, single-valued function connecting light curve shape
and color.

\subsection{Luminosity}

\begin{figure}[!h]
\epsscale{0.7}
\vspace{-1.0cm}
\plotone {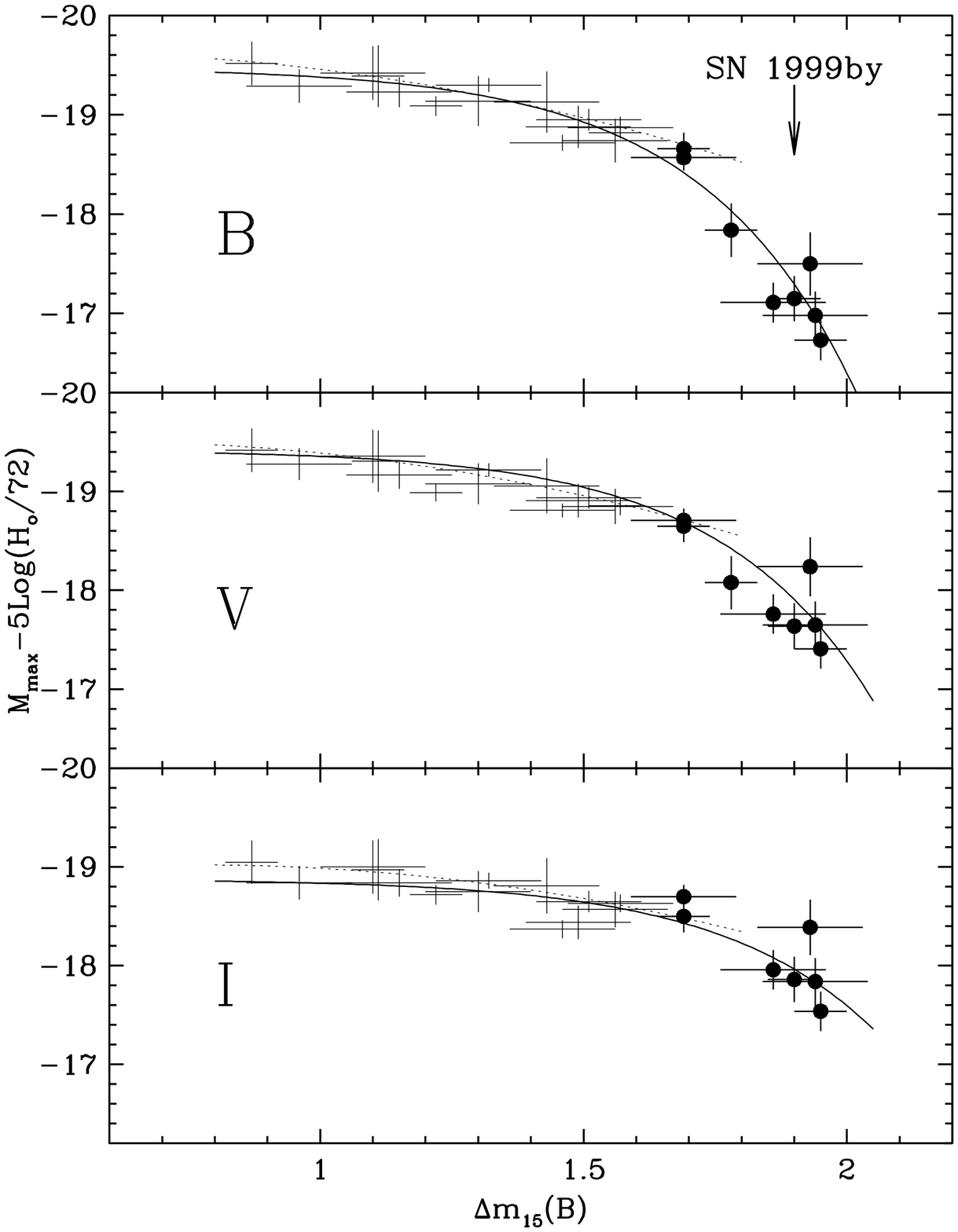}
\caption{The absolute magnitudes of SNe Ia versus \dm15\ from
Phillips et al. (1999), with so-called ``peculiar'' supernovae added (solid 
points). The dotted line is the quadratic fit derived by Phillips et al. for 
\dm15$<1.7$. The solid line is an exponential fit which attempts to fit
all the objects represented.}
\label{fig:absm}
\end{figure}

Several ``peculiar'' events listed in Table~6 are distant enough to be in the
Hubble flow so their luminosities can be compared with so-called normal SNe Ia.  
Figure~\ref{fig:absm} shows the absolute magnitudes derived for the fast
decliners vs. \dm15, assuming $H_0=72$~\kms$\;$Mpc$^{-1}$ (Freedman et al. 2001).  
The photometric error bars include an uncertainty of 400~\kms\ in the recession
velocity due to the unknown peculiar velocities of the hosts.  SN~1999by is
included by using the Cepheid distance of 14.1$\pm 1.5$~Mpc to NGC~2841 from
Macri et al. (2001). The SNe Ia with smaller decline rates are from Phillips et
al. (1999) and all magnitudes have been corrected using the reddening estimates
from that work. The $B$-band luminosities show a steep decline toward high \dm15\
which is not well fit by an extension of the Phillips et al. (1999) quadratic
fit. Instead, we use an exponential function to match decline rate with
luminosity. The best fits for $B$, $V$ and $I$ are

\begin{equation}
M(B)=-19.338+5{\rm Log}(H_0/72)+0.139\; \biggl({\rm exp}[3.441\; (\Delta m_{15}(B)-1.1)]-1\biggr)
\end{equation}
\begin{equation}
M(V)=-19.328+5{\rm Log}(H_0/72)+0.096\; \biggl({\rm exp}[3.450\; (\Delta m_{15}(B)-1.1)]-1\biggr)
\end{equation}
\begin{equation}
M(I)=-18.817+5{\rm Log}(H_0/72)+0.060\; \biggl({\rm exp}[3.402\; (\Delta m_{15}(B)-1.1)]-1\biggr)
\end{equation}

\parindent 0 mm

using a $\chi^2$ minimization with three free parameters. Adding an additional
linear slope parameter did not improve the overall fit which has an
rms scatter of 0.19~mag in each band. While this is a larger
scatter than obtained when fitting SNe Ia restricted to \dm15$<1.7$, it shows
that so-called ``normal'' and ``peculiar'' SNe Ia can be calibrated in a uniform
manner, which argues for their common origin.

\parindent 9 mm

However, while there is evidence that the absolute magnitudes in the $UBVRI$
bands present a monotonic sequence, SN~1999by is unusual regarding its IR maxima.  
Krisciunas, Phillips, \& Suntzeff (2004a) show that SNe Ia appear to have $JHK$
absolute magnitudes at maximum which are independent of the decline rate
parameter, at least for the range of 0.8 $\lesssim$ \dm15 $\lesssim$ 1.7.  We
obtain M$_H$(max) = $-$17.87 $\pm$ 0.24 and M$_K$(max) = $-$17.79 $\pm$ 0.25.  
These are to be compared with the mean values of $-$18.25 and $-$18.42,
respectively, for the objects analyzed by Krisciunas et al. (2004a).  SN~1999by is
$\Delta$H = 0.38 $\pm$ 0.24 and $\Delta$K = 0.63 $\pm$ 0.25 mag fainter than the
mean absolute magnitudes of the slower decliners.  While both of these values
are less than 3-$\sigma$ outliers, it is noteworthy that SN~1999by has the
faintest IR absolute magnitudes at maximum yet found for SNe Ia.  It could be
that the fast-declining SNe Ia have statistically different absolute magnitudes
at maximum in the IR compared to other SNe Ia.
%\footnote[12]{We have recently
%obtained well-sampled optical and IR imagery of SN~2003gs, a fast decliner with
%\dm15 = 1.84 (Suntzeff et al. 2005). At the time of this writing we have not yet
%determined if its IR absolute magnitudes are close to the mean values found by
%Krisciunas et al. (2004a) or similar to those of SN~1999by.}

\section{Two for One: SN~1957A}

The rather prolific galaxy NGC~2841 has also hosted SN~1957A, which was identified as
a subluminous Type~I supernova by Branch \& Doggett (1985). Branch, Fisher, \&
Nugent (1993) describe it as a SN~1991bg-like event and a reanalysis of its
photographic spectra by Casebeer et al. (2000) shows a flux deficit around 420~nm
relative to ``normal' SNe Ia at a similar age, which suggests the presence of
Ti~II.  However, in the blue part of the spectrum a Type~Ic supernova can be
mistaken for subluminous SN Ia after maximum, so SN~1957A can not be classified 
as a Type~Ia with absolute certainty.  Still, we can analyze the historical light
curve compiled by Leibundgut et al. (1991) as if it was a Type~Ia.

We fit the blue photographic light curve by stretching the Leibundgut template
and find that maximum light occurred on Julian Day 2,435,901$\pm 2$ at 
$m_{pg}=14.43\pm
0.2$. The best fit stretch parameter is 0.54$\pm 0.03$ corresponding to a
\dm15$=2.00\pm 0.07$. The $V$-band light curve is not as well sampled, so we use
the same stretch factor and time of maximum to find a $V_{max}=13.63\pm 0.2$. The
color of SN~1957A between 30 and 90 days from maximum is not very well determined
given the errors on faint photographic magnitudes, but we will assume minimal
extinction from the host. We convert between standard $B$-band and $m_{pg}$ using
equation~31 from Pierce \& Jacoby (1995) and find SN~1957A was 0.88 mag fainter
than SN~1999by in $B$ and 0.48 mag fainter in $V$. If SN~1957A was truly an
unreddened SN Ia, then it was the faintest yet observed with $M_B=-16.3\pm 0.2$
and $M_V=-17.2\pm 0.2$. These are consistent with the luminosities predicted by
equations~2 and 3 for \dm15$=2.00$ of $M_B=-16.40$ and $M_V=-17.28$. The
intrinsic color we find from the light curves of SN~1957A is
$B_{max}-V_{max}=0.87$ which is close to the 0.81 mag expected from its decline
rate (see Figure~\ref{fig:dm15color}).

\section{The Hubble Constant}

The measurement of a Cepheid distance to NGC~2841 provides an opportunity
to make an independent estimate of the Hubble parameter using fast-declining SNe 
Ia alone. Jha et al. (1999) noted that most of the SNe Ia 
calibrated with Cepheids have slow light curve decline rates. If there is a 
systematic bias caused by this selection, then the Hubble parameter estimated 
from fast decliners may be significantly different from the Key Project
value (Freedman et al. 2001). However, with only one directly calibrated fast 
decliner the uncertainty on the derived Hubble parameter can be no better than 10 
percent.

\begin{figure}[!h]
\epsscale{0.7}
\vspace{-1.0cm}
\plotone {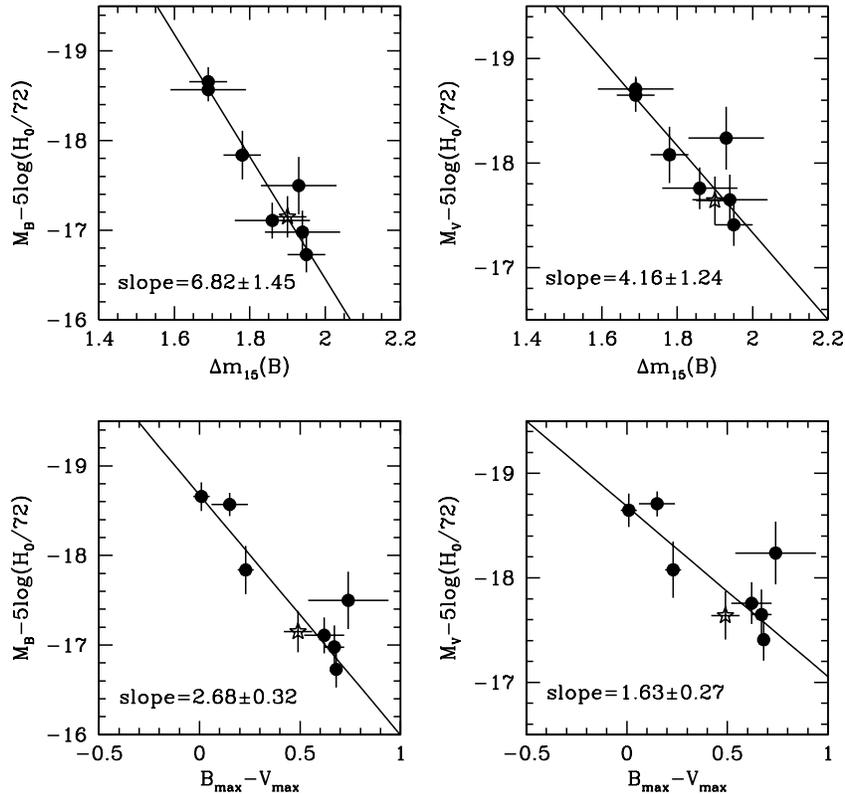}
\caption{The absolute magnitudes of fast-declining SNe Ia versus \dm15\ and
$B_{max}-V_{max}$. The star indicates the values for SN~1999by although it
is not used in the linear fit.}
\label{fig:absm_pec}
\end{figure}

There are seven SNe Ia listed in Table~6 that can be considered in the Hubble
flow, i.e. with recession velocities in excess of 3000~\kms. We plot the
luminosities of these in Figure~\ref{fig:absm_pec} (assuming
$H_0=72$~\kms$\;$Mpc$^{-1}$) as a function of \dm15. We also plot the
luminosities versus $B_{max}-V_{max}$ because it appears to be highly correlated
with the light curve shape in the fast decliners. The slope of the linear fit to
the points (excluding SN~1999by) is then used to estimate a magnitude correction
to the fiducial values \dm15=1.90 and $B_{max}-V_{max}=0.49$ (extinction
corrected).  SN~1992K is consistently the most discrepant point from the linear
trend and its errors in decline rate and peak brightness may be underestimated.

Using the method described in Jha et al. (1999), we derive a Hubble constant of
$H_0=75^{+12}_{-11}$~\kms$\;$Mpc$^{-1}$ applying the \dm15\ parameter to correct
the seven events to SN~1999by. Using $B_{max}-V_{max}$ as the luminosity
indicator, we find $H_0=84^{+12}_{-11}$~\kms$\;$Mpc$^{-1}$. These are consistent
with the Key Project value derived from normal SNe Ia (Freedman et al. 2001) and
suggest that no systematic error in the Hubble constant is being introduced by
ignoring low luminosity events.

\section{Conclusions}

We present well-sampled $UBVRI$ light curves of SN~1999by and the evolution of
the spectra over a period of 40~days around maximum. We estimate a light curve
parameter \dm15=1.90$\pm 0.05$ mag which is one of the fastest observed decline
rates for a SN Ia. The observed color at maximum $B_{max}-V_{max}=0.51$ mag,
which is red for a SN Ia, but the late-time $B-V$ color suggests little dust extinction
along the line-of-sight. From the recent Cepheid distance to the host galaxy
(Macri et al. 2001) and the assumption of minimal host extinction, we find the
absolute brightness at maximum to be $M_B=-17.15\pm 0.23$ and $M_V=-17.64\pm
0.23$.

To date SN~1999by has the faintest absolute magnitudes at maximum in the
near-IR of any known Type Ia supernova.  Whether SN~1999by is a statistical 
outlier or this facet is representative of other fast decliners is not yet known.
While most SNe Ia may be considered IR standard candles (Meikle 2000, Krisciunas
et al. 2004a), the objects with \dm15 $\gtrsim$ 1.7 might be subluminous.

Comparing SN~1999by with other so-called ``peculiar'' SNe Ia, we find that it had
a slower decline rate and was not as red at maximum as SN~1991bg or SN~1998de. It
was slightly more luminous than SN~1998de, but fainter than SN~1998bp. There
appears to be an excellent correlation between luminosity, decline rate and color
for these subluminous objects and these observables blend smoothly into the
luminous population of SNe Ia. We show that any SN Ia with a decline rate of
\dm15$<2.0$ can be used as a distance indicator, although the scatter increases
with large \dm15\ due to the steep correction curve and poor calibration at the
faint end of the population. We test this by using the fast decliners to estimate
the Hubble constant and find it consistent with the value found from the
``normal'' population.
  
We find that the 580~nm feature commonly associated with Si~II and correlated
with luminosity is actually dominated by Ti~II for the SN~1991bg-like events and
is likely to have a significant Ti~II component even for more standard SNe Ia.  
This means that the depth ratio of 580~nm to 615~nm should not be used as a
luminosity indicator for \dm15$<1.2$, but it still makes an excellent 
temperature
and luminosity estimator for supernovae with faster than normal light curves.
Predictions of the Ti~II strength with temperature suggest that Ti~II optical
depth increases rapidly over a small range of temperature. It is this non-linear
effect that makes SN~1991bg-like events appear so odd when compared to more
typical SNe Ia which change their character little over a wide temperature span.  
We conclude that SN~1991bg-like supernovae should not be considered as
``peculiar'' SNe Ia since along with ``Branch normal'' events they form a
continuous, smooth, and single-valued distribution of luminosity, color, Ti~II
strength, and perhaps even $^{56}$Ni yield.

\acknowledgements

We thank the FLWO observers who helped gather these data: P. Berlind, N.
Caldwell, M. Calkins, K. Dendy, E. Falco, S. Kenyon, J. Mader, T. Megeath, D.  
McIntosh, M. Pahre, K. Rines, K. Stanek, A. Szentgyorgyi, A. Tustin, and P.
Vaisanen.  The authors also thank T. Matheson for useful suggestions and
discussions. We are grateful to the Vatican Observatory Research Group and R.
Boyle for generous allotments of VATT nights. We also thank the NSF supported REU
program at the Harvard-Smithsonian Center for Astrophysics. PMG was partially
supported by NASA LTSA grant NAG5-9364 and RPK acknowledges support from NSF
grant AST98-19825.  We gratefully made use of the NASA/IPAC Extragalactic
Database (NED) and the Two Micron All Sky Survey (2MASS).

%\clearpage

%\clearpage

\begin{deluxetable}{ccccccc}
\tablecolumns{7}
\tablewidth{0pc}
\tablecaption{Local Standard Star Magnitudes}
\startdata
\tableline \tableline Star & $U$ & $B$ & $V$ & $R$ & $I$ & n\tablenotemark{a} \\ \tableline
1 & 14.642 & 14.262 & 13.503 & 13.059 & 12.656 & 3 \\
2 & 15.490 & 14.999 & 14.185 & 13.725 & 13.312 & 4 \\
3 & 16.131 & 16.191 & 15.708 & 15.407 & 15.088 & 6 \\
4 & 17.748 & 17.335 & 16.530 & 16.049 & 15.583 & 6 \\
5 & 16.948 & 16.182 & 15.269 & 14.713 & 14.209 & 2 \\
6 & 16.842 & 15.713 & 14.650 & 13.994 & 13.501 & 2 \\
7 & 18.630 & 18.332 & 17.575 & 17.127 & 16.701 & 6 \\ 
8 & 18.931 & 18.207 & 17.314 & 16.798 & 16.298 & 6 \\ 
9 & 18.333 & 18.101 & 17.363 & 16.904 & 16.472 & 6 \\
10 & 18.598 & 17.946 & 17.049 & 16.488 & 16.085 & 4 \\
\enddata
\tablenotetext{a}{Number of nights calibrated.}
\end{deluxetable}

\begin{deluxetable}{ccccccl}
\tablecolumns{7}
%\tabletypesize{\scriptsize}
%\rotate
\tablewidth{0pc}
\tablecaption{$UBVRI$ Photometry of SN 1999by\tablenotemark{a}}
\startdata
\tableline \tableline JD\tablenotemark{b}  &  $U$  & $B$   &   $V$   &  $R$  &  $I$ 
&  Observer \\ \tableline 
1304.67  &  13.96 (02) &  13.95 (02) &  13.64 (02) &  13.49 (02) &  13.40 (02) & Caldwell\\ 
1305.68  &  13.89 (02) &  13.81 (02) &  13.49 (02) &  13.33 (02) &  13.26 (02) & Caldwell\\ 
1306.67  &  13.84 (02) &  13.74 (02) &  13.37 (02) &  13.20 (02) &  13.14 (02) & Rines\\ 
1307.64  &  13.82 (02) &  13.67 (02) &  13.27 (02) &  13.09 (02) &  13.05 (02) & Rines\\    
1308.63  &  13.84 (02) &  13.66 (02) &  13.21 (02) &  13.02 (02) &  12.99 (02) & Rines\\   
1309.63  &  13.89 (02) &  13.68 (02) &  13.17 (02) &  12.97 (02) &  12.94 (02) & Rines\\    
1310.73  &  13.99 (02) &  13.72 (02) &  13.14 (02) &  12.94 (02) &  12.91 (02) & McIntosh\\    
1311.64  &  14.12 (03) &  13.80 (02) &  13.14 (02) &  12.93 (02) &  12.91 (02) & McIntosh\\    
1312.70  &  14.30 (03) &  13.92 (02) &  13.17 (02) &  12.95 (02) &  12.90 (02) & Garnavich\\   
1313.66  &  14.46 (03) &  14.07 (02) &  13.23 (02) &  12.98 (02) &  12.93 (02) & Dendy\\    
1314.65  &  14.67 (03) &  14.24 (02) &  13.31 (02) &  13.03 (02) &  12.95 (02) & Dendy\\   
1315.68  &  14.84 (03) &  14.43 (02) &  13.42 (02) &  13.10 (02) &  12.99 (02) & Dendy\\   
1316.65  &  15.06 (04) &  14.61 (02) &  13.52 (02) &  13.17 (02) &  13.01 (02) & Pahre\\   
1318.65  &  15.38 (05) &  14.97 (02) &  13.76 (02) &  13.33 (02) &  13.08 (02) & Pahre\\   
1335.69  &  16.50 (07) &  16.21 (03) &  15.08 (03) &  14.65 (03) &  14.19 (03) & Szentgyorgyi\\   
1338.66  &  16.57 (07) &  16.33 (03) &  15.21 (03) &  14.81 (03) &  14.36 (03) & Szentgyorgyi\\    
1340.66  &  16.63 (07) &  16.40 (03) &  15.29 (03) &  14.93 (03) &  14.48 (03) & Szentgyorgyi\\    
1342.67  &  16.71 (10) &  16.45 (05) &  15.37 (05) &  15.03 (06) &  14.59 (06) & Szentgyorgyi\\    
1348.66  &  16.68 (15) &  16.60 (08) &  15.59 (06) &  15.35 (09) &  14.91 (09) & Falco\\    
1350.65  &  16.77 (07) &  16.66 (04) &  15.65 (03) &  15.42 (04) &  15.00 (04) & Falco\\
1487.97  &  20.67 (30) &  19.72 (05) &  19.62 (05) &  19.61 (05) &  18.80 (05) & Stanek\\
1521.99  &    \nodata  &  20.39 (20) &  20.17 (12) &  20.38 (20) &  19.45 (30) & Garnavich\\
%1529.11  &   \nodata  &   \nodata   &  20.36 (05) &   \nodata   &    \nodata  & Jha\\
%1581.88  &    \nodata &  21.19 (14) &  21.29 (14) &  21.67 (28) &  20.39 (17) & Falco\\
\enddata
\tablenotetext{a}{The values in parentheses are the uncertainties in hundredths
of a magnitude.}
\tablenotetext{b}{Julian Date $-$ 2,450,000.}
\end{deluxetable}

%\clearpage

\begin{deluxetable}{ccccl}
\tablecolumns{5}
\tablewidth{0pc}
\tablecaption{Infrared Photometry of SN 1999by}
\startdata
\tableline \tableline JD\tablenotemark{a}  &  $J$   &  $H$  &  $K$  & Observer \\ \tableline 
%1299.80 &  \nodata       & 14.650 (0.088) &   \nodata      & Tustin \\
%1301.63 & 14.164 (0.046) & 14.156 (0.090) & 14.196 (0.082) & Vaisanen \\
%1303.63 & 13.439 (0.031) & 13.329 (0.070) &   \nodata      & Vaisanen \\
%1320.71 & 13.679 (0.027) & 13.022 (0.033) & 13.213 (0.044) & Pahre \\
%1322.71 & 13.710 (0.029) & 13.179 (0.036) & 13.255 (0.049) & Pahre \\
%1324.64 & 13.820 (0.034) & 13.335 (0.036) & 13.312 (0.041) & Kenyon \\
%1326.63 & 13.875 (0.034) & 13.512 (0.067) & 13.676 (0.073) & Kenyon \\
%1327.67 & 14.010 (0.035) & 13.607 (0.073) & 13.777 (0.052) & Garnavich \\
%1328.64 & 14.202 (0.052) & 13.739 (0.075) & 13.723 (0.076) & Garnavich \\
%1330.67 & 14.231 (0.054) & 14.068 (0.076) & 14.124 (0.079) & Megeath \\
%1360.64 &   \nodata      & 15.737 (0.172) &   \nodata      & Mader  \\
%
1299.80 &  \nodata     & 14.65 (0.09) &   \nodata      & Tustin \\
1301.63 & 14.16 (0.05) & 14.16 (0.09) & 14.20 (0.08) & Vaisanen \\
1303.63 & 13.44 (0.03) & 13.33 (0.07) &   \nodata      & Vaisanen \\
1320.71 & 13.68 (0.03) & 13.02 (0.03) & 13.21 (0.04) & Pahre \\
1322.71 & 13.71 (0.03) & 13.18 (0.04) & 13.25 (0.05) & Pahre \\
1324.64 & 13.82 (0.03) & 13.33 (0.04) & 13.31 (0.04) & Kenyon \\
1326.63 & 13.87 (0.03) & 13.51 (0.07) & 13.68 (0.07) & Kenyon \\
1327.67 & 14.01 (0.04) & 13.61 (0.07) & 13.78 (0.05) & Garnavich \\
1328.64 & 14.20 (0.05) & 13.74 (0.08) & 13.72 (0.08) & Garnavich \\
1330.67 & 14.23 (0.05) & 14.07 (0.08) & 14.12 (0.08) & Megeath \\
1360.64 &   \nodata    & 15.74 (0.17) &   \nodata      & Mader  \\
\enddata
\tablenotetext{a} {Julian Date $-$ 2,450,000.}
\end{deluxetable}

\begin{planotable}{lcccl}
\tablecaption{Log of Spectroscopic Observations}
\tablewidth{15cm}
\tablehead{
\colhead{UT Date\tablenotemark{a}} &
\colhead{ JD\tablenotemark{b} } &
\colhead{ Coverage (nm) } &
\colhead{ Exposure Time (s)} &
\colhead{ Observer } 
}
\startdata
May 6  & 1304.72 & 362-754 & $3\times 300$ & Berlind \nl
May 7  & 1305.63 & 362-754 & $2\times 660$ & Calkins\nl
May 8  & 1306.64 & 327-901 & $2\times 480$ & Calkins\nl
May 9  & 1307.64 & 362-560 & $600$ & Dendy\nl
May 13 & 1311.67 & 362-560 & $600$ & Dendy\nl
May 14 & 1312.64 & 327-940 & $3\times 480$ & Berlind\nl
May 15 & 1313.65 & 362-754 & $2\times 360$ & Berlind\nl
May 16 & 1314.67 & 362-754 & $2\times 420$ & Berlind\nl
May 17 & 1315.66 & 362-754 & $3\times 300$ & Garnavich\nl
May 18 & 1316.63 & 362-754 & $3\times 420$ & Garnavich\nl
May 19 & 1317.63 & 362-754 & $3\times 480$ & Garnavich\nl
May 21 & 1319.64 & 500-750 & $2\times 600$ & Calkins\nl
May 22 & 1320.64 & 390-595 & $2\times 600$ & Calkins\nl
Jun 5  & 1334.66 & 362-754 & $2\times 600$ & Calkins\nl
Jun 9  & 1338.66 & 362-754 & $2\times 600$ & Berlind\nl
Jun 11 & 1340.65 & 362-754 & $900$ & Calkins\nl
Jun 13 & 1342.65 & 362-754 & $900$ & Calkins\nl
Jun 22 & 1351.65 & 362-754 & $900$ & Calkins\nl
\enddata
\tablenotetext{a} {Year = 1999.}
\tablenotetext{a} {Julian Date $-$ 2,450,000.}
\end{planotable}

\begin{planotable}{ccc}
\tablewidth{10cm}
\tablecaption{Light Curve Parameters at Maximum Light}
\tablehead{
\colhead{Band} &
\colhead{JD$_{max}$\tablenotemark{a}} &
\colhead{ Maximum magnitude\tablenotemark{b} }
}
\startdata
$U$ & 1307.6 (0.3) & 13.82 (0.03) \nl
$B$ & 1308.8 (0.3) & 13.66 (0.02) \nl
$V$ & 1310.8 (0.4) & 13.15 (0.02) \nl
$R$ & 1311.4 (0.4) & 12.94 (0.02) \nl
$I$ & 1311.8 (0.5) & 12.91 (0.03) \nl
$H$ & 1314.0 (2.0) & 12.89 (0.04) \nl
$K$ & 1313.5 (2.0) & 12.96 (0.09) \nl
\enddata
\tablenotetext{a} {Julian Date $-$ 2,450,000.}
\tablenotetext{b} {These are observed values, uncorrected 
for extinction along the line of sight.}
\end{planotable}

\begin{planotable}{lcccccl}
\tablecaption{SN~Ia with Fast-Declining Light Curves\tablenotemark{a}}
\tablehead{
\colhead{ SN } &
\colhead{ $\Delta$m$_{15}$($B$)} &
\colhead{ cz } &
\colhead{ $B_0$(max) } &
\colhead{ $V_0$(max) } &
\colhead{ $I_0$(max) } &
\colhead{ Reference }
}
\startdata
1992bo & 1.69 (05) & 5445\tablenotemark{b} & 15.73 (07) & 15.75 (06) & 15.90 (05) & Hamuy et al. (1996) \nl
1993H  & 1.69 (10) & 7447\tablenotemark{b} & 16.51 (08) & 16.37 (05) & 16.38 (06) & Hamuy et al. (1996) \nl
1986G  & 1.73 (07) & 547  &  9.90 (30) &  9.46 (30)  & \nodata    & Phillips et al. (1987; 1999) \nl
1998bp & 1.78 (05) & 3127\tablenotemark{b} & 15.29 (07) & 15.06 (05) & \nodata    & Jha et al. (2001) \nl
1997cn & 1.86 (10) & 5246\tablenotemark{b} & 17.20 (10) & 16.55 (10) & 16.35 (10) & Turatto et al. (1998) \nl
1999by & 1.90 (05) & 638  & 13.59 (03) & 13.10 (03) & 12.88 (03) & This work \nl
1992K  & 1.93 (10) & 3334\tablenotemark{b} & 15.83 (21) & 15.09 (16) & 14.94 (15) & Hamuy et al. (1996) \nl
1991bg & 1.93 (05) & 1060 & 14.58 (05) & 13.82 (03) & 13.50 (05) & Leibundgut et al. (1993)\nl
1999da & 1.94 (10) & 3748\tablenotemark{b} & 16.60 (06) & 15.93 (06) & 15.74 (04) & Krisciunas et al. (2001) \nl
1998de & 1.95 (05) & 4625\tablenotemark{b} & 17.31 (04) & 16.63 (04) & 16.50 (05) & Modjaz et al. (2001) \nl
1957A\tablenotemark{c}   & 2.00 (07) & 638  & 14.47 (20) & 13.58 (20)  & \nodata  & This work \nl
\enddata
\tablenotetext{a}{Radial velocities given in column 3 are measures in \kms.  Magnitudes in
columns 4, 5, and 6 are corrected for extinction along the line of sight.}
\tablenotetext{b}{Corrected to CMB frame.}
\tablenotetext{c}{Probable Type Ia supernova.}
\end{planotable}

\end{document}